\documentclass{iau-JDSS}
\usepackage{graphicx}

\newcommand{\oversim}[2]{\protect{\mbox{\lower0.5ex\vbox{%
  \baselineskip=0pt\lineskip=0.2ex
  \ialign{$\mathsurround=0pt #1\hfil##\hfil$\crcr#2\crcr\sim\crcr}}}}}
\newcommand{\simless} {\mbox{$\,\mathrel{\mathpalette\oversim<}\,$}} 

\title[short title of paper] 
{
The high-mass stellar IMF in different environments
}

\author[Pavel Kroupa]   
{Pavel Kroupa}

\affiliation{
Argelander Institute for Astronomy, University of Bonn, Auf dem H\"ugel 71, D-53121 Bonn, Germany
\break email: pavel@astro.uni-bonn.de \\[\affilskip]
}

\pubyear{2006}
\volume{Volume 14}  
\pagerange{119--126}
\date{?? and in revised form ??}
\jname{Highlights of Astronomy, Volume 14}
\editors{K.A. van der Hucht, ed.}
\begin{document}

\maketitle

\vspace{-1mm}

\begin{abstract}
The massive-star IMF is found to be invariable. However, integrated
IMFs probably depend on galactic mass.  
\end{abstract}


\noindent
To set the stage it is useful to emphasise that the
\cite{Salpeter1955} mass function (MF), $\xi(m) = k\,m^{-\alpha},
\alpha=2.35$, is strictly valid only for stellar masses with
$0.4\simless m/M_\odot \simless 10$. The number of stars in the mass
interval $m,m+dm$ is $dN = \xi(m)\,dm$. Given statistical noise in the
IMF of otherwise equal systems the question we need to answer is
whether there is any significant empirical evidence for systematic
variation of the IMF with the physical conditions of star
formation. This question is of fundamental importance for
star-formation theory because an observed systematic variation poses
constraints on the theory. The question is also of fundamental
importance for cosmology because the physical conditions of star
formation have changed dramatically over a cosmological epoch implying
possible systematic changes of the young stellar populations with
cosmic time and therefore systematic changes in the properties of
galaxies.  Very different theoretical approaches lead to one common
result, namely that the average stellar mass ought to shift towards
larger values with decreasing metallicity.  Thus,
\cite{AdamsFatuzzo1996} develop a model of the origin of the IMF based
on the notion that stars regulate their own masses through feedback,
while \cite{Larson1998} investigates the systematic changes of the IMF
as a result of the temperature-dependence of the Jeans mass. Recent
developments based on a change of the equation of state as a result of
dust processes may be an avenue of explaining the general absence of a
variation of the stellar mass at which the IMF peaks,
\cite{Bonnell_etal2006}.  Observationally, the IMF has been
constrained above a few~$M_\odot$ by \cite{Massey2003} for populations
in the Milky Way, the Large and Small Magellanic Clouds and for
different densities. The massive-star IMF has been found to have an
invariable ``Salpeter/Massey'' index $\alpha = 2.3\pm0.2$. This
insensitivity to the physical conditions needs to be understood, and
perhaps suggests that massive star formation is dominated by
scale-free processes such as coagulation and/or competitive accretion
in the very dense environments where massive stars form,
\cite{Bonnell_etal2006}. For entire galaxies the composite IMF
probably depends on galaxy type such that low-mass galaxies have very
steep indices, \cite{WeidnerKroupa2006}, despite the invariability of
the IMF.

\end{document}